\documentclass[%
 reprint,
superscriptaddress,
 amsmath,amssymb,
 aps,
]{revtex4-2}

\usepackage{graphicx}% Include figure files
\usepackage{float}
\usepackage{graphicx}
\usepackage{dcolumn}% Align table columns on decimal point
\usepackage{bm}% bold math
\usepackage{verbatim}
\usepackage[dvipsnames]{xcolor}
\usepackage{hyperref}% add hypertext capabilities
\usepackage[mathlines]{lineno}% Enable numbering of text and display math
\usepackage{makecell}
\usepackage{comment}
\begin{document}

\preprint{APS/123-QED}

\title{Entanglement-based quantum key distribution with data in hollow-core fiber}
%\title{Coexistence of Time-encoded High-dimensional Entanglement Key and Classical Data over Hollow-Core Fiber}

\author{Yue Luo}
\affiliation{Institute of Fundamental and Frontier Sciences, University of Electronic Science and Technology of China, Chengdu 611731, China}
\affiliation{Center for Quantum Internet, Tianfu Jiangxi Laboratory, Chengdu 641419, China}
\affiliation{Key Laboratory of Quantum Physics and Photonic Quantum Information, Ministry of Education, University of Electronic Science and Technology of China, Chengdu 611731, China}

\author{Sheng Liu}
\affiliation{Department of Fundamental Network Technology, China Mobile Research Institute, Beijing 100053, China}

\author{Yun-Ru Fan}
\email{yunrufan@uestc.edu.cn}
\affiliation{Institute of Fundamental and Frontier Sciences, University of Electronic Science and Technology of China, Chengdu 611731, China}
\affiliation{Center for Quantum Internet, Tianfu Jiangxi Laboratory, Chengdu 641419, China}
\affiliation{Key Laboratory of Quantum Physics and Photonic Quantum Information, Ministry of Education, University of Electronic Science and Technology of China, Chengdu 611731, China}

\author{Da-Wei Ge}
\affiliation{Department of Fundamental Network Technology, China Mobile Research Institute, Beijing 100053, China}

\author{Zhi-Yang Liu}
\affiliation{Institute of Fundamental and Frontier Sciences, University of Electronic Science and Technology of China, Chengdu 611731, China}
\affiliation{Center for Quantum Internet, Tianfu Jiangxi Laboratory, Chengdu 641419, China}
\affiliation{Key Laboratory of Quantum Physics and Photonic Quantum Information, Ministry of Education, University of Electronic Science and Technology of China, Chengdu 611731, China}

\author{Hao Li}
\affiliation{Shanghai Institute of Microsystem and Information Technology, Chinese Academy of Sciences, Shanghai 200050, China}

\author{Si Shen}
\affiliation{Southwest Institute of Technical Physics, Chengdu 610041, China}

\author{Zi-Chang Zhang}
\affiliation{Southwest Institute of Technical Physics, Chengdu 610041, China}

\author{Hai-Zhi Song}
\affiliation{Institute of Fundamental and Frontier Sciences, University of Electronic Science and Technology of China, Chengdu 611731, China}
\affiliation{Southwest Institute of Technical Physics, Chengdu 610041, China}

\author{Li-Xing You}
\affiliation{Shanghai Institute of Microsystem and Information Technology, Chinese Academy of Sciences, Shanghai 200050, China}

\author{Tao Zhou}
\affiliation{School of Automation Engineering, University of Electronic Science and Technology of China, Chengdu 611731, China}

\author{Kai Guo}
\email{guokai07203@hotmail.com}
\affiliation{Institute of Systems Engineering, AMS, Beijing 100141, China}

\author{Guang-Can Guo}
\affiliation{Institute of Fundamental and Frontier Sciences, University of Electronic Science and Technology of China, Chengdu 611731, China}
\affiliation{Center for Quantum Internet, Tianfu Jiangxi Laboratory, Chengdu 641419, China}
\affiliation{Key Laboratory of Quantum Physics and Photonic Quantum Information, Ministry of Education, University of Electronic Science and Technology of China, Chengdu 611731, China}
\affiliation{CAS Center for Excellence in Quantum Information and Quantum Physics, University of Science and Technology of China, Hefei 230026, China}

\author{Qiang Zhou}
\email{zhouqiang@uestc.edu.cn}
\affiliation{Institute of Fundamental and Frontier Sciences, University of Electronic Science and Technology of China, Chengdu 611731, China}
\affiliation{Center for Quantum Internet, Tianfu Jiangxi Laboratory, Chengdu 641419, China}
\affiliation{Key Laboratory of Quantum Physics and Photonic Quantum Information, Ministry of Education, University of Electronic Science and Technology of China, Chengdu 611731, China}
\affiliation{CAS Center for Excellence in Quantum Information and Quantum Physics, University of Science and Technology of China, Hefei 230026, China}

\date{\today}
\begin{abstract}
The coexistence of quantum information and classical signals in a single fiber is essential for future quantum networks that leverage the well-established optical fiber infrastructure. Although multiplexing technologies can separate quantum and classical signals, pure silica core fibers (PSCFs) remain fundamentally limited by the high nonlinearity, which generates substantial Raman scattering and four-wave mixing noise. Hollow-core fibers (HCFs), guiding light predominantly in air, offer an attractive solution with intrinsically ultra-low nonlinearity and strongly suppressed nonlinear noise. In this work, we demonstrate the entanglement-based key coexisting with data over an 18-km HCF link. We achieve time-encoded high-dimensional quantum key distribution (HD-QKD) carrying 0~dBm of bidirectional received power, corresponding to a theoretical data capacity of up to 2.3~Tbps. During 24~hours of continuous operation, an average secret key rate (SKR) of 10.56~kbps is obtained. Theoretical analysis further predicts SKRs above 135~kbps over transmission distances exceeding 200~km using state-of-the-art low-loss HCFs. These results show significantly improved performance compared with PSCF-based systems and highlight the potential of HCFs for scalable quantum–classical coexistence compatible with the architectures of established fiber-optic networks.
\end{abstract} 

\maketitle
\section{\label{sec:level1}Introduction}
As quantum communication technologies progress toward practical implementation, integrating them into existing optical network infrastructure has become increasingly attractive\cite{wehner2018quantum, valencia2026large}. To facilitate this integration, several multiplexing techniques—wavelength-division multiplexing (WDM)\cite{thomas2024quantum, rahmouni2024100, zy2d-m3ch}, time-division multiplexing (TDM)\cite{patel2012coexistence, wang2024time}, and space-division multiplexing (SDM)\cite{kong2023coexistence, yu2026coexistence}—have been investigated to enable the coexistence of quantum and classical signals within a single optical fiber. However, most demonstrations of quantum-classical coexistence have relied on pure silica core fibers (PSCFs), particularly conventional single-mode fibers (SMFs). In such fibers, the overlap between the optical field and the silica material leads to strong nonlinear noise when high-power classical signals co-propagate with weak quantum signals. Consequently, quantum signals become highly susceptible to nonlinear noise sources, including spontaneous Raman scattering (SpRS) and four-wave mixing (FWM)\cite{ferreira2014impact, Dou:24, dibaj2024traffic}, which fundamentally limit the performance of quantum protocols.

Hollow-core fibers (HCFs) provide a promising alternative platform for quantum-classical coexistence. Guided by the Fabry–Pérot anti-resonant mechanism formed by thin-walled capillaries\cite{numkam2023loss}, HCFs drastically reduce the interaction between the optical field and the silica material, resulting in nonlinear coefficients several orders of magnitude lower than those of PSCFs\cite{honz2023first,liu2025quantum}. In addition, HCFs exhibit substantially lower attenuation and dispersion than conventional SMFs, with losses as low as 0.04~dB/km and dispersion of 2–3~$\text{ps}/(\text{nm}\cdot\text{km})$\cite{li2026low, gao202540}, compared to $\sim$0.14~dB/km and $\sim$17~$\text{ps}/(\text{nm}\cdot\text{km})$ in state-of-the-art SMFs\cite{sato2024record}. Recent studies have demonstrated the coexistence of quantum key distribution and classical data transmission in HCFs, highlighting their potential for low-noise coexistence system\cite{nasti2022utilizing, alia2022dv, honz2023first, clark2025coexistence, kong2025experimental}. However, realizing sufficiently high secret key rates (SKRs) for practical secure communication remains a central challenge, requiring further improvements in system performance.

\begin{figure*}[htb]
    \centering
    \includegraphics[width=18 cm]{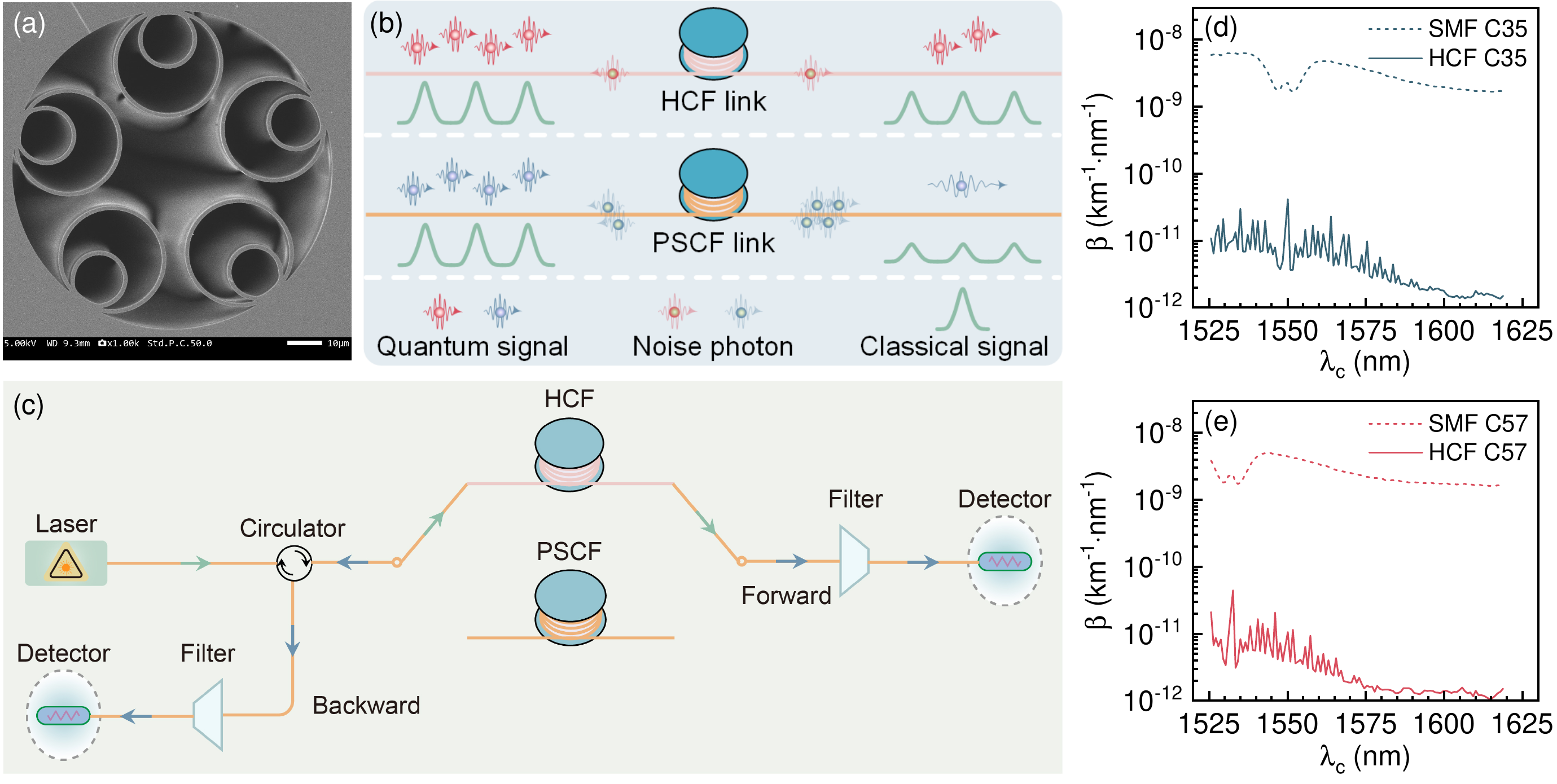}
    \caption{Performance comparison between SMF and HCF for quantum-classical coexistence. (a) SEM image of HCF in this work. (b) Conceptual illustration of quantum-classical coexistence performance in SMF and HCF transmission systems. (c) Schematic of the Raman scattering measurement principle. (d) Measured Raman scattering coefficients of SMF (dashed) and HCF (solid) in ITU channels C35 (blue) and (e) C57 (red) with different $\lambda_c$.}
    \label{fig:Fig1}
\end{figure*}

In this Letter, we demonstrate the coexistence of time-encoded high-dimensional quantum key distribution (HD-QKD) and classical data in HCF. Our experimental system employs an 18-km HCF link to transmit entanglement-based key in the C bands, while bidirectional tunable classical channels operate across the C and L bands with a received power of 0~dBm. This configuration corresponds to a theoretical classical data capacity of up to 2.3~Tbps, as estimated using Shannon’s formula\cite{poggiolini2022opportunities}. By achieving polarization management and continuous-window-based time-sifting scheme, an average SKR of 10.56~kbps is obtained during 24~hours of continuous operation, with quantum bit error rates (QBERs) of 16.00\% and 15.86\% in the time and phase bases, respectively. Theoretical analysis based on state-of-the-art low-loss HCFs indicates that an SKR exceeding 135~kbps can be maintained over transmission distances exceeding 200~km. These results establish HCFs as a viable platform for large-scale quantum-classical coexistence in future optical networks.

\section{Properties of HCF}
As shown in the scanning electron microscope (SEM) image in Fig.~{\ref{fig:Fig1}}(a), HCFs predominantly guide light in an air core\cite{ding2014analytic, wang2017confinement}. This dramatically reduces light–matter interaction, resulting in ultra-low nonlinearity, while their intrinsically low dispersion and reduced attenuation further enhance transmission performance.

The HCF used in this work was provided by the China Mobile Research Institute, with a measured attenuation coefficient of $0.27~\mathrm{km}^{-1}$. The total coupling loss for the 18-km link is approximately 2.5 dB. The measured transmission loss shows no significant dependence on the input polarization.

As illustrated in the quantum-classical coexistence concept in Fig.~{\ref{fig:Fig1}}(b), these properties allow HCFs to suppress noise generation, preserve the temporal structure of both quantum and classical signals, and improve overall transmission efficiency, making them a highly promising platform for coexistence systems\cite{kimble2008quantum, wehner2018quantum}.

In contrast, conventional PSCFs confine light within a silica core, where high-power classical channels induce strong nonlinear scattering. This contaminates the quantum channel, leading to excess QBER and severely limiting the achievable SKR and transmission distance.

To quantify nonlinear noise, we experimentally measure the forward Raman scattering coefficients in both HCF and SMF at two quantum-channel wavelengths\cite{eraerds2010quantum, clark2025coexistence, patel2012coexistence}, which are at $\lambda_s = 1549.32$ nm (ITU C35) and $\lambda_i = 1531.90$ nm (ITU C57). Here, the Raman scattering coefficient is defined as the Raman noise power generated per unit optical bandwidth and per unit fiber length, normalized by the launched classical signal power. The detailed expression is provided in Supplementary Material Section~\uppercase\expandafter{\romannumeral 1}. Figure~{\ref{fig:Fig1}}(c) presents a simplified schematic of the Raman scattering measurement principle. A continuous-wave laser is launched into the fiber through a circulator. The forward Raman noise generated in the fiber is filtered out at the fiber output using an optical filter, while the backward Raman noise is separated by the circulator and extracted using another filter. The filtered Raman photons are detected by single-photon detectors, and their count rates are recorded. 

\begin{figure*}[htb]
    \centering
    \includegraphics[width=18 cm]{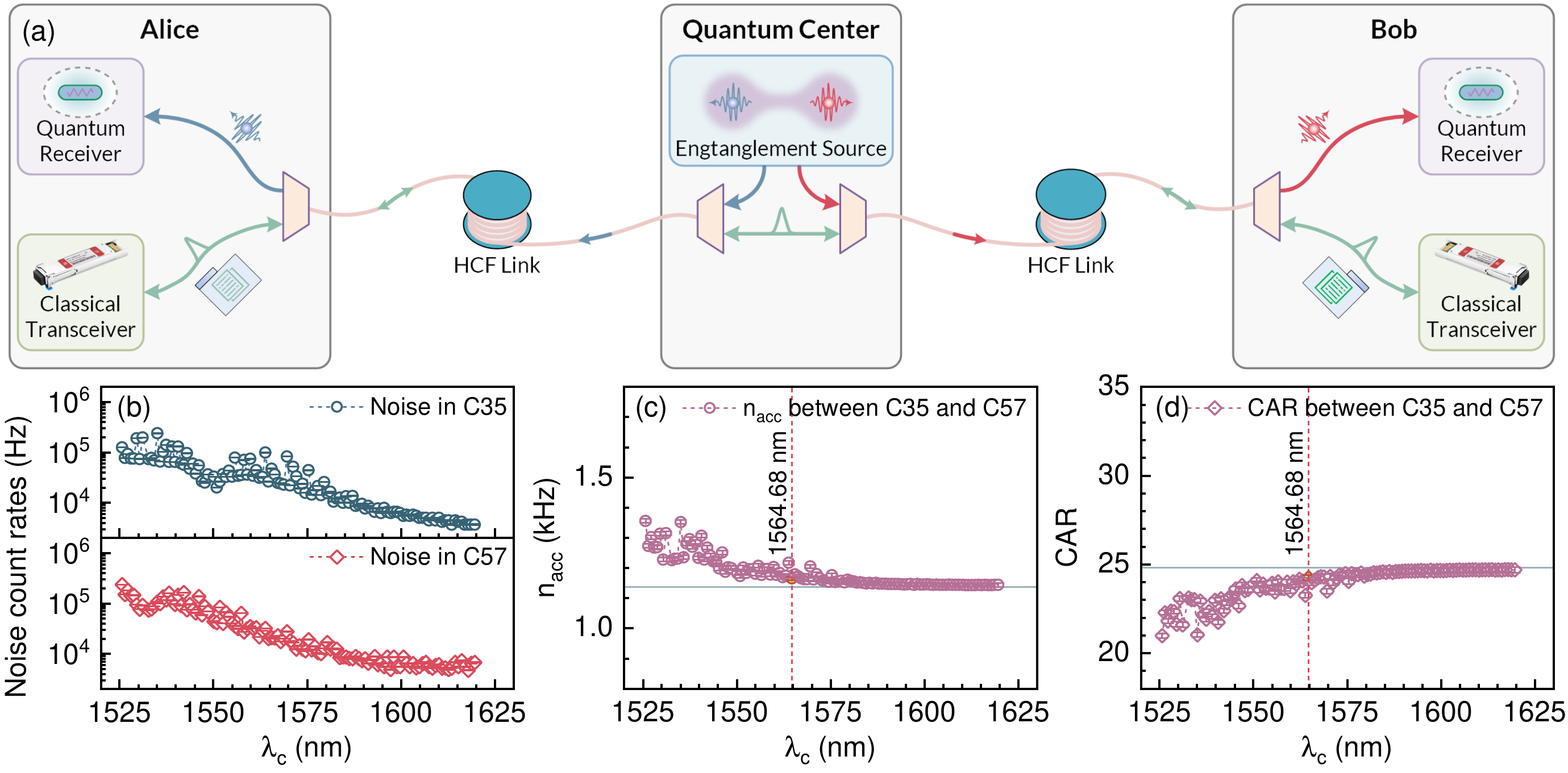}
    \caption{Characterization of HCF-based coexistence system. (a) Schematic of system noise measurement. (b) Single-side count rates of noise photons in ITU channels C35 (blue) and C57 (red) with different $\lambda_c$. (c) Accidental coincidence count rate and (d) CAR of noise photons between ITU channels C35 and C57 with different $\lambda_c$.}
    \label{fig:Fig2}
\end{figure*}

Based on the measured count rates, together with the input optical power and fiber attenuation coefficient, the forward and backward Raman scattering coefficients are calculated. More details of the complete experimental setup and principle are provided in Supplementary Material Section~\uppercase\expandafter{\romannumeral 1}. As shown in Figs.~{\ref{fig:Fig1}}(d) and (e), the forward Raman scattering coefficient in HCF is reduced by two to three orders of magnitude compared with SMF, indicating a substantial suppression of nonlinear noise generation.

The advantages in nonlinear noise, attenuation, and dispersion enables significantly higher launched classical powers under coexistence conditions while maintaining low QBERs. This directly translates into extended transmission distances and enhanced SKRs in HD-QKD-based coexistence systems.

\section{HD-QKD under quantum-classical coexistence}
Two 9-km segments of HCF are employed in our experiment. Figure~{\ref{fig:Fig2}}(a) illustrates the experimental principle for simultaneous quantum and classical signal transmission. Alice and Bob are connected through the HCF link, while a quantum center located between them distributes energy-time entangled photon pairs generated by the entanglement source to both users, where they are detected by quantum receivers. In parallel, Alice and Bob each employ a classical transceiver to send and receive classical communication signals. The classical and quantum signals are multiplexed into different wavelength and polarization channels to suppress inter-channel crosstalk and noise. The detailed experimental setup is provided in Supplementary Material Section~\uppercase\expandafter{\romannumeral 2}.

To characterize noise originating from the classical channels, we measure the single-side count rates of noise photons in two quantum channels under classical transmission. The classical wavelength $\lambda_c$ is scanned from 1525 to 1625~nm with a received power of -15~dBm, while the entanglement source and the phase-basis measurement modules are disconnected. During bidirectional classical data transmission between Alice and Bob, polarization controllers (PC1--PC4) are adjusted to minimize the single-side count rates in the quantum channels. The results are shown in Fig.~{\ref{fig:Fig2}}(b), corresponding to the power spectral density of noise induced by the classical data channels.

We then reconnect the entanglement source and evaluate the impact of Raman noise on quantum correlations. With a fixed pump power of 22~mW and a coincidence time window of 300~ps, the accidental coincidence count rate between photon pairs in C35 and C57 channels are measured, as shown in Fig.~{\ref{fig:Fig2}}(c). Based on the measured accidental and total coincidence counts, the coincidence-to-accidental ratio (CAR) is subsequently calculated, as presented in Fig.~\ref{fig:Fig2}(d). Within the C band, the lowest accidental coincidence count rate and the highest CAR are observed when $\lambda_c$ is set to 1564.68~nm, indicating the lowest noise level. This wavelength is selected as the operating wavelength of the classical channel for subsequent experiments. And the classical launch power is amplified to achieve a received power of 0~dBm, corresponding to a theoretical classical capacity of up to 2.3~Tbps, as estimated using Shannon’s formula\cite{poggiolini2022opportunities}.

\begin{figure*}[htb]
    \centering
    \includegraphics[width=18 cm]{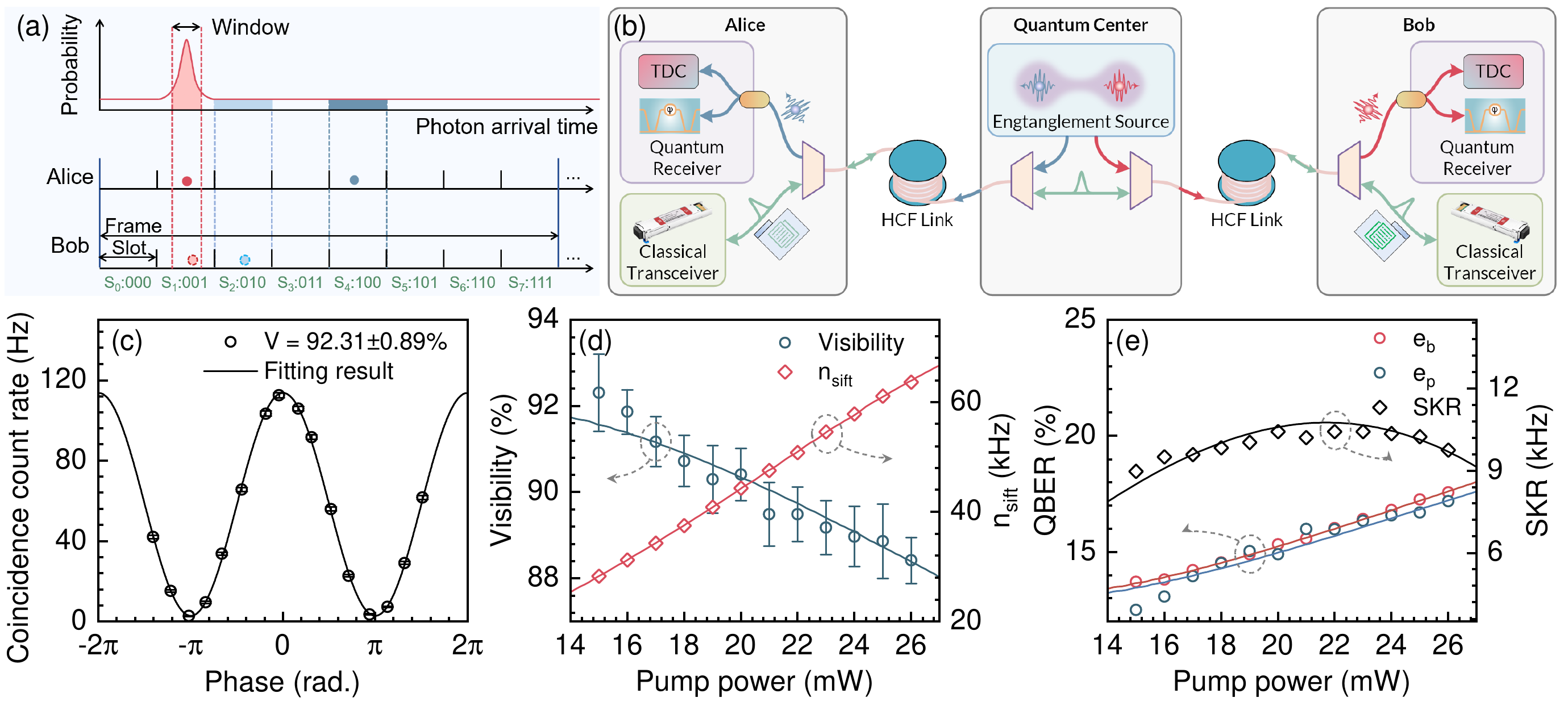}
    \caption{Experimental demonstration of HD-QKD under quantum-classical coexistence conditions. (a) Conceptual illustration of the protocol, where red and blue balls denote entangled and noise photons, respectively. (b) Schematic of the HD-QKD system. (c) Franson interference curve at a pump power of 15~mW. (d) Visibilities and sifted key rates under different pump powers. (e) QBERs and SKRs under different pump powers.}
    \label{fig:Fig3}
\end{figure*}

To evaluate the key distribution performance of the system under quantum-classical coexistence conditions, we implement a continuous-window-based three-level time-sifting HD-QKD protocol using energy-time entangled photon pairs distributed to Alice and Bob\cite{liu2019energy, liu2024high}. Figure~\ref{fig:Fig3}(a) conceptually illustrates the protocol, where red and blue balls denote entangled and noise photons, respectively. The photon arrival times recorded at Alice and Bob are divided into consecutive frames with equal duration. Each frame is further partitioned into multiple slots, and raw keys are generated according to the slot indices in which the photons arrive. To further exploit the temporal correlation properties of energy-time entangled photon pairs and improve the SKR, we employ a continuous-window scheme instead of the conventional discrete-bin scheme in the third-level time-sifting process, which is conceptually similar to the mode-pairing strategy\cite{zeng2022mode, zhu2023experimental}. Detailed descriptions of both schemes are provided in Supplementary Material Section~\uppercase\expandafter{\romannumeral 3}.

Figure~\ref{fig:Fig3}(b) shows the schematic of the HD-QKD system. The phase-basis measurement modules are reinserted, and the entangled photon pairs are randomly directed into time- and phase-basis measurement modules via a beam splitter with an 80:20~splitting ratio. The time-encoded dimension is set to $d = 8$, corresponding to 3~bits per frame. The slot width is set to 1250~ps, which is determined by the path-length difference of the interferometers\cite{yu2025quantum}. The continuous-window width $T_w$ is set to 85~ps.

To characterize the phase-basis performance, the phase at Alice is scanned while keeping the phase at Bob fixed, thereby obtaining Franson interference curves from the three-level time-sifting procedure. Figure~\ref{fig:Fig3}(c) shows the measured Franson interference curve at a pump power of 15~mW, yielding a visibility of 92.31 $\pm$ 0.89\%. We further investigate the dependence of system performance on pump power. As shown in Fig.~\ref{fig:Fig3}(d), the visibility gradually decreases with increasing pump power due to enhanced noise, but remains above 88\% throughout the measured range. Meanwhile, the sifted key rate continuously increases because more entangled photon pairs are generated and detected.

The degradation of visibility is mainly attributed to two factors. First, the Raman noise generated within the entangled photon-pair source increases with pump power, reducing the signal-to-noise ratio of the detected photon pairs. Second, increasing the pump power also raises the probability of generating multiple entangled photon pairs within the same frame. Since at most one coincidence event can be effectively registered from a single frame, these multi-pair emissions reduce the temporal correlation between the detected photon pairs, thereby degrading the measured interference visibility. In contrast, the Raman noise induced by the coexisting classical signals remains nearly constant during these measurements because the received classical power is fixed.

Based on the measured visibility and coincidence statistics, the QBERs and SKRs are calculated. During the experiment, 1\% of the time-basis measurement outcomes, together with all phase-basis measurement outcomes, are used to estimate the QBERs in time and phase bases. The remaining time-basis outcomes are used for secret key generation according to\cite{zhong2024hyperentanglement,yu2025quantum}
\begin{equation}
    \text{SKR} \geq \frac{n_{\text{sift}}}{\log_2d}[\log_2d-f(e_b)H_d(e_b)-H_d(e_p)]\,
\end{equation}
where $n_{\text{sift}}$ represents the sifted key rate, $d$ is time-encoded dimension, $f(x)$ is the error correction efficiency as a function of error rate (here we set it to 1.2), $H_d(x)$ is the $d$-dimensional Shannon entropy function defined as $H_d(x)=-x\log_2[x/(d-1)]-(1-x)\log_2(1-x)$, and $e_b$ and $e_p$ represent the QBERs in time and phase bases, respectively. $e_p$ is estimated from the visibility $V$ of the Franson interference curve as\cite{yu2025quantum}
\begin{equation}
    e_b = \frac{1}{2}-\frac{V}{d-V(d-2)}\,
\end{equation}

\begin{figure*}[htb]
    \centering
    \includegraphics[width=18 cm]{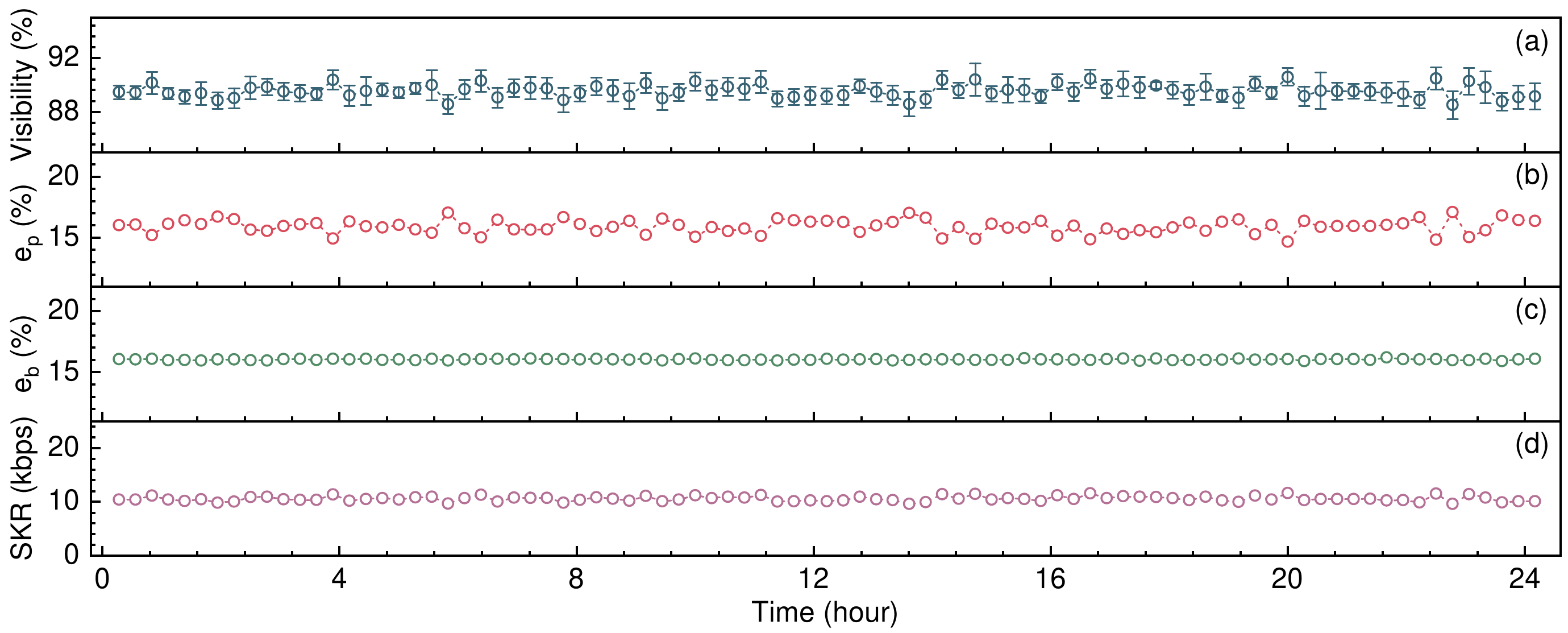}
    \caption{Long-term stability test of the HD-QKD system over 24~hours under quantum-classical coexistence conditions. (a) Visibility. (b) $e_p$. (c) $e_b$. (d) SKR.}
    \label{fig:Fig4}
\end{figure*}

Figure~\ref{fig:Fig3}(e) presents the measured QBERs and SKRs under different pump powers. As the pump power increases, the SKR first rises due to the enhanced pair generation rate and then gradually saturates because of increased noise contributions. The maximum SKR of 10.43~kbps is achieved at a pump power of 22~mW, corresponding to the optimal operating point. At this pump power, the measured QBERs in time and phase bases are 16.41\% and 16.34\%, respectively, both remaining well below the theoretical threshold of 21.70\% for 8-dimensional HD-QKD.

\begin{table}[!ht]
\centering
\caption{Comparison of HD-QKD performance for 18-km HCF and loss-equivalent B2B transmission with and without coexisting classical signals.}
\begin{tabular}{cccccc}
\hline
\textbf{Case} & \textbf{\begin{tabular}[c]{@{}c@{}}Classical\\Signal\end{tabular}} & \textbf{\begin{tabular}[c]{@{}c@{}}Visibility\\(\%)\end{tabular}} & \textbf{\begin{tabular}[c]{@{}c@{}}$e_b$\\(\%)\end{tabular}} & \textbf{\begin{tabular}[c]{@{}c@{}}$e_p$\\(\%)\end{tabular}} & \textbf{\begin{tabular}[c]{@{}c@{}}SKR\\(kbps)\end{tabular}}\\
\hline
18 km HCF & Off & 92.68 $\pm$ 0.74 & 11.60 & 12.00 & 21.01\\
B2B + VOA & Off & 93.73 $\pm$ 0.60 & 10.82 & 10.55 & 21.88\\
18 km HCF & On & 89.49 $\pm$ 0.63 & 16.03 & 15.98 & 10.43\\
B2B + VOA & On & 94.24 $\pm$ 0.80 & 9.77 & 9.82 & 19.82\\
\hline
\end{tabular}
\label{tab:comparison}
\end{table}

To further distinguish the effects of fiber transmission from those of coexisting classical signals, we performed a comparative experiment under four conditions: 18-km HCF and back-to-back (B2B) + VOA transmission, each with and without classical signals. In the B2B configuration, variable optical attenuators (VOAs) were used to reproduce the transmission loss of the 18-km HCF link, thereby providing a loss-equivalent reference with negligible fiber-induced nonlinear effects. The B2B measurements were performed using the corresponding optimized operating parameters determined for the 18-km HCF measurements. The measured visibility, time-basis QBER, phase-basis QBER, and SKR are summarized in Table~\ref{tab:comparison}.

In the absence of classical signals, the B2B + VOA and HCF configurations exhibited comparable visibility, QBERs, and SKRs, indicating that the intrinsic impact of the HCF on HD-QKD performance is relatively small under equivalent transmission loss. When classical signals were introduced, the B2B + VOA configuration maintained substantially better quantum-channel performance than the HCF configuration, as evidenced by its higher visibility, lower QBERs, and higher SKR. These results indicate that the HCF link exhibits significantly greater performance degradation under the investigated quantum-classical coexistence conditions than the corresponding loss-equivalent B2B reference.

Note that the lower SKR of the B2B + VOA coexistence case than its corresponding no-classical case, despite its lower QBERs, is mainly due to the smaller time-window parameter $T_w$ selected for noise suppression in the HCF coexistence case and subsequently used in the corresponding B2B measurement. Although this tighter time window suppresses noise, it also reduces the probability of retaining valid detection events, thereby decreasing the sifted-key rate $n_{\mathrm{sift}}$ and resulting in a lower SKR.

To evaluate the long-term operational stability of the system, the HD-QKD protocol is continuously operated for 24~hours under quantum-classical coexistence conditions with the pump power fixed at the optimal value of 22~mW. The measured visibility, phase-basis QBER, time-basis QBER, and SKR are shown in Figs.~\ref{fig:Fig4}(a)-(d), respectively. Each data point corresponds to a measurement duration of 1000~s. In the asymptotic regime, an average SKR of 10.56~kbps is achieved. The measured average QBERs in time and phase bases are 16.00\% and 15.86\%, respectively. The stable visibility, consistently low QBERs, and positive SKR throughout the 24-hour operation demonstrate the robustness and long-term stability of the proposed quantum-classical coexistence system based on HCF.

\begin{figure*}[htb]
    \centering
    \includegraphics[width=18 cm]{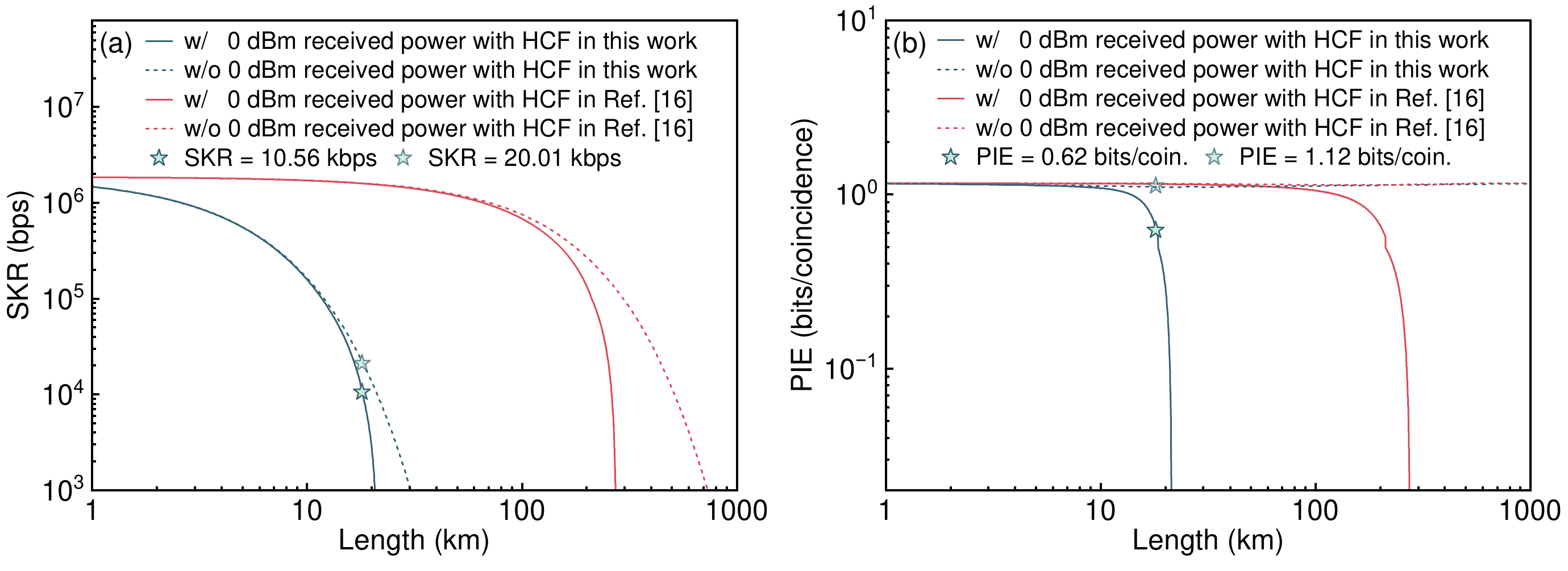}
    \caption{Simulated HD-QKD performance on the fiber link length. (a) SKR. (b) PIE. Curves are shown for the simulated results of our HCF (blue) and HCF with the lowest reported attenuation\cite{li2026low} (red), with (solid) and without (dashed) classical signals. Experimental measurement results are indicated by stars.}
    \label{fig:Fig5}
\end{figure*}

\begin{table*}[!ht]
\centering
\caption{A summary of entanglement-based quantum key distribution coexisting with data.}
\begin{tabular*}{18cm}{@{\extracolsep{\fill}}cccccccc}
\hline
\textbf{Reference} & \textbf{\begin{tabular}[c]{@{}c@{}}Fiber\\Type\end{tabular}} & \textbf{\begin{tabular}[c]{@{}c@{}}Length\\(km)\end{tabular}} & \textbf{\begin{tabular}[c]{@{}c@{}}Received Power\\(dBm)\end{tabular}} & \textbf{\boldmath $d(\log_2 d)^a$} & \textbf{\begin{tabular}[c]{@{}c@{}}Visibility\\(\%)\end{tabular}} & \textbf{\begin{tabular}[c]{@{}c@{}}QBER\\(\%)\end{tabular}} & \textbf{\begin{tabular}[c]{@{}c@{}}SKR\\(kbps)\end{tabular}}\\
\hline
This work & HCF & 18 & 0 & 8(3) & 89.17 $\pm$ 0.63 & 16.00 & 10.56\\
\cite{clark2025coexistence} & HCF & 11.5 & -17.46$^b$ & 2(1) & 90.0 $\pm$ 0.8 & 5.4 & 0.005$^c$ \\
\cite{fan2023energy} & SMF & 40 & -27 & 2(1) & 84.85 $\pm$ 2.07 & 8.61 & 0.343 \\
\cite{jing2024coexistence} & SMF & 45.6 & -2.32$^d$ & – & 74 & – & – \\
\cite{rahmouni2024100} & SMF & 100 & -41 & – & 81 $\pm$ 2 & – & – \\
\cite{zhong2024hyperentanglement} & SMF & 50 & -19.2 & 4+2(2+1)$^e$ & – & – & 0.007 \\
\cite{thomas2024quantum} & SMF & 30.2 & -18 & – & 95.5 $\pm$ 0.2 & – & – \\
\cite{luo2025quantum} & SMF & 40 & -24 & 2(1) & 84.54 $\pm$ 0.72 & 8.09 & 1.89 \\
\hline
 & HCF & 100 & 10 & 4(2) & 86.12 & 12.19 & 189.7\\
 & HCF & 100 & 0 & 8(3) & 93.05 & 11.50 & 681.2\\
Theoretical$^f$ & HCF & 200 & 0 & 8(3) & 90.25 & 15.08 & 134.7\\
 & SMF & 40 & -24 & 8(3) & 91.68 & 13.32 & 235.9\\
 & SMF & 65 & -24 & 4(2) & 81.55 & 15.58 & 2.498\\
\hline
\end{tabular*}
\label{tab:tableI}

\raggedright
$^a$The entries with $d=2$ correspond to implementations based on the Bennett-Brassard-Mermin 1992 (BBM92) protocol\cite{bennett1992quantum}. $^b$ This power corresponds to the output power at the fiber end, excluding the insertion losses of subsequent filtering and isolation components. $^c$ The data correspond to the best-performing coexistence link reported in this referenced work. $^d$ The received power is estimated assuming a fiber loss of 0.2~dB/km. $^e$ In this referenced work, time and polarization are encoded in 4 and 2 dimensions, respectively. $^f$ These theoretical results are obtained for different received-power and transmission-length configurations selected from Supplementary Material Section~\uppercase\expandafter{\romannumeral 7} assuming state-of-the-art low-loss HCFs reported in~\cite{li2026low}.
\end{table*}

To assess the performance limits of HCF-based quantum-classical coexistence under the HD-QKD protocol, we perform a theoretical analysis of the SKR as a function of the fiber link length, assuming a fixed classical received power of 0~dBm. The theoretical framework is provided in Supplementary Material Section~\uppercase\expandafter{\romannumeral 5}. The characterization results of the entanglement source used in the simulation are provided in Supplementary Material Section~\uppercase\expandafter{\romannumeral 6}. In the simulations, the slot width is fixed at 1250~ps, while the time-encoded dimension $d$, pump power, and continuous-window width $T_w$ are jointly optimized at each distance to maximize the SKR. The simulation results are shown in Fig.~{\ref{fig:Fig5}}(a). In addition, we also evaluate the corresponding photon information efficiency (PIE)\cite{zhong2015photon}, defined as $\text{PIE}=\log_2d\cdot SKR/n_{\text{sift}}$, which quantifies the number of secret bits extracted per coincidence event. The calculated PIE is presented in Fig.~\ref{fig:Fig5}(b).

For an attenuation coefficient of $\alpha = 0.27~\text{km}^{\text{-1}}$, corresponding to the HCF used in the present experiment, the SKR and PIE decreases rapidly with increasing transmission distance. This decay is primarily attributed to the relatively high attenuation assumed in the simulations. 

It is worth noting that, under the fixed received classical power adopted in this work (0 dBm), reducing the fiber attenuation does not increase the Raman noise despite the longer effective interaction length. Instead, a lower attenuation requires a lower launched classical power to maintain the same received power, resulting in a lower Raman noise level according to Eqs. (S2) and (S3) in Supplementary Material Section~\uppercase\expandafter{\romannumeral 7}. Specifically, if the attenuation were reduced to the level of SMF, the Raman noise would decrease to approximately one-sixth of that measured in this work, while using state-of-the-art low-loss HCFs would further reduce the Raman noise to approximately one-seventh.

Furthermore, the reduced attenuation allows significantly more entangled photon pairs to reach the receivers, thereby improving the signal-to-noise ratio and enhancing the robustness of HD-QKD against classical-noise-induced errors. In contrast, when HCFs with the lowest reported attenuation are considered\cite{li2026low}, the analysis indicates that an SKR exceeding 135~kbps can be maintained over transmission distances beyond 200~km.

For comparison, we also simulate the performance of an otherwise identical system implemented with conventional SMF. In this case, the achievable SKR is fundamentally limited by stronger nonlinear noise, resulting a substantially reduced performance limit compared with HCF-based systems. A detailed comparison between HCF and SMF is provided in Supplementary Material Section~\uppercase\expandafter{\romannumeral 7}.

Table~\ref{tab:tableI} summarizes recent demonstrations of entanglement distribution coexisting with classical data transmission. Compared with previous experimental studies in both HCFs and SMFs, our work achieves the highest reported SKR under coexistence conditions, reaching 10.56~kbps at a received classical power of 0~dBm while employing an 8-dimensional encoding scheme. In contrast, most previous demonstrations operated at significantly lower received powers and achieved SKRs ranging from a few bps to a few kbps. Theoretical analysis further indicates that, with state-of-the-art low-loss HCFs, high-dimensional entanglement distribution could maintain high visibility and SKR over 100–200~km links, substantially outperforming comparable SMF-based configurations. These results highlight the potential of combining HCF transmission and high-dimensional encoding for scalable quantum–classical coexistence networks.

\section{Discussion and conclusion}
In conclusion, we experimentally demonstrate the coexistence of entanglement-based key and classical data transmission over a HCF link. The system uses a continuous-window-based three-level time-sifting scheme to further exploit the correlation properties of the entangled photon pairs and improve the SKR. Using WDM techniques and polarization management, quantum entanglement is successfully distributed over an 18-km HCF carrying bidirectional classical signals with a received power of 0~dBm. Under continuous operation for 24~hours, an average SKR of 10.56~kbps is achieved using a three-level time-sifting HD-QKD protocol, with measured QBERs of 16.00\% and 15.86\% in time and phase bases, respectively. Theoretical analysis based on state-of-the-art low-loss HCFs indicates that an SKR exceeding 135~kbps can be maintained over transmission distances exceeding 200~km.

Toward practical deployment of quantum-classical coexistence systems, further advances are required at both the system and protocol levels. On the system side, higher-brightness and hyperentanglement  sources\cite{zhong2024hyperentanglement}, together with lower-loss and vacuum-core HCFs\cite{michieletto2016hollow}, would enable higher SKR and extended transmission distances while further suppressing nonlinear noise. In realistic communication networks employing multiple wavelength channels, joint optimization of quantum and classical channel allocation will also be necessary to mitigate four-wave mixing noise\cite{niu2018optimized, wu2025integration}. In addition, active stabilization techniques, including automatic polarization control and adaptive power and wavelength management, will be essential for robust operation in dynamically varying network environments\cite{shen2023hertz, gavignet2023co}. On the protocol side, while our demonstration is analyzed in the asymptotic regime, incorporating finite-key security analysis remains an important step toward accurately assessing performance under realistic operating conditions\cite{liu2024high, niu2016finite}.

\begin{acknowledgments}
This work was supported by Sichuan Science and Technology Program (Nos. 2024YFHZ0368, 2025YFHZ0337, 2024YFHZ0369, 2024YFHZ0370), National Natural Science Foundation of China (Nos. 62405046, 62475039), Quantum Science and Technology-National Science and Technology Major Project (Nos. 2024ZD0300800, 2021ZD0300701), and Tianfu Jiangxi Laboratory (No. TFJX-ZD-2025-005).

YF and QZ conceived and supervised the project. YL mainly carried out the experiment and collected the experimental data with help of other authors. SL and GD provided HCFs used in the experiment. LY developed and maintained the SNSPDs used in the experiment. YL analyzed the data. YL and YF wrote the manuscript with inputs from all other authors. All authors have given approval for the final version of the manuscript.
\end{acknowledgments}

\section*{Data Availability}
There are no publicly available research data or software supporting this manuscript. Requests for further information or data should be sent to the authors.

\appendix

\section{Materials and Methods}
\subsection{Residual gas composition in the hollow core}
During fabrication, HCF and its preform was filled with high-purity $N_2$ gas, and after fabrication, ambient air was pushed into HCF as the as-drawn HCF maintain a sub-atmosphere pressure of 0.2 bar. Hence, trace gases like $O_2$, $CO_2$ and water vapor appear in HCF, while $N_2$ is the dominate gas component. Since the optical mode is predominantly confined within the hollow core, the measured Raman spectra exhibit characteristic Raman peaks originating from the residual gases inside the hollow core. A quantitative evaluation of the respective Raman-noise contributions from the residual gases and the surrounding silica films is beyond the scope of this work and would require dedicated measurements, such as replacing the gas inside the hollow core with different gases or evacuating the fiber. The overall Raman noise in HCF is nevertheless substantially lower than that in conventional silica-core fibers because the overlap between the guided optical mode and silica is dramatically reduced.

\subsection{Attenuation and Raman spectra characterizations}
A broadband light source with a wavelength range from 1525 nm to 1620 nm and a total output power of 10 dBm was used together with an OSA with a resolution of 0.02 nm to characterize the attenuation spectra of the HCF and SMF. For Raman spectra characterizations, a tunable continuous-wave (CW) laser was scanned from 196.5 THz (1525.661 nm) to 185.1 THz (1619.624 nm) with a step of 0.1 THz. During the Raman spectra measurements, the optical power launched into the fiber was monitored using a power meter to calculate the Raman scattering coefficients of the HCF and SMF.

%\bibliography{reference}


\end{document}